\documentclass[matprg]{svjour}

\usepackage[dvips]{graphicx}
\usepackage{numinsec}
\usepackage{url}
\usepackage{times}

\DeclareSymbolFont{AMSb}{U}{msb}{m}{n}
\DeclareMathSymbol{\R}{\mathbin}{AMSb}{"52}
\newcommand{\cP} {\mbox{$\cal P$}}
\newcommand{\cS} {\mbox{$\cal S$}}

\newcommand{\Ref}[1]{\mbox{\rm{(\ref{#1})}}}
\newcommand{\lt}{<}

\newcommand{\cops} {\mbox{\sf \footnotesize COPS}}

\begin{document}
\title{Benchmarking Optimization Software with Performance Profiles}

\author{Elizabeth D. Dolan \and Jorge J. Mor\'e}

\institute{Elizabeth D. Dolan \at Northwestern University and 
Mathematics and Computer Science Division, Argonne National Laboratory. 
\email{dolan@mcs.anl.gov} \and Jorge J. Mor\'e \at
Mathematics and Computer Science Division, Argonne National Laboratory.
\email{more@mcs.anl.gov}}

\combirunning{E.D.Dolan \and J.J.Mor\'e: Performance Profiles}

\maketitle 

\begin{abstract}
We propose performance profiles --- distribution 
functions for a performance metric~--- as
a tool for benchmarking and comparing optimization software. 
We show that performance profiles combine the best features
of other tools for performance evaluation.


\end{abstract}

\section{Introduction}

\label{intro}
The benchmarking of optimization software has recently gained
considerable visibility.   Hans Mittlemann's
\cite{Benchmarks} work on a variety of optimization software
has frequently uncovered deficiencies in the software and has
generally led to software improvements.
Although Mittelmann's efforts have gained the most notice,
other researchers have been concerned with the evaluation and
performance of optimization codes. 
As recent examples, we cite 
\cite{HYB00,SCB95a,ASB98,BCGT97,CGT96,HM99,RJV99}.

The interpretation and analysis of the data generated by the
benchmarking process are the main technical issues addressed
in this paper.  Most benchmarking efforts involve tables
displaying the performance of each solver on each problem for a set
of metrics such as CPU time, number of function evaluations, or
iteration counts for algorithms where an iteration implies a
comparable amount of work.  Failure to display such tables for a small
test set would be a gross omission, but they tend
to be overwhelming for large test sets. In all cases, the interpretation
of the results from these tables is often a source of
disagreement.

The quantities of data that result
from benchmarking with large test sets have spurred researchers to try
various tools for analyzing the data. 
The solver's average or cumulative total for
each performance metric over all problems is sometimes 
used to evaluate performance \cite{HYB00,BCGT97,CGT96}.
As a result, a small number of the most difficult problems
can tend to dominate these results, and researchers must take pains to
give additional information.  Another drawback is that computing
averages or totals for a performance metric 
necessitates discarding problems for which any
solver failed, effectively biasing the results
against the most robust solvers.  As an alternative to disregarding
some of the problems, a penalty value can be assigned for failed
solver attempts, but this requires a subjective choice for 
the penalty. Most researchers choose to report the
number of failures only in a footnote or separate table.

To address the shortcomings of the previous approach, some researchers
rank the solvers 
\cite{BCGT97,CGT96,SGN91,RJV99}. In other words, they count the
number of times that a solver comes in $k$th place, usually for $ k =
1, 2 , 3 $.  Ranking the solvers' performance for each problem helps
prevent a minority of the problems from unduly influencing the
results.
Information on the size of the improvement, however, is lost.

Comparing the medians and quartiles of some performance metric
(for example, the difference between solver times \cite{BCGT97})
appears to be a viable way of ensuring that 
a minority of the problems do not dominate the results,
but in our testing we have witnessed large leaps in quartile values
of a performance metric, rather than gradual trends. 
If only quartile data is used, then information on
trends occurring between one quartile and the next is lost; and we
must assume that the journey from one point to another proceeds at a
moderate pace.  
Also, in the specific case of contrasting the differences
between solver times, the comparison fails to provide any
information on the relative size of the improvement.
A final drawback is that if results are mixed, interpreting
quartile data may be no easier than using the raw data;
and dealing with comparisons of more than two solvers might become 
unwieldy.

The idea of comparing solvers by the ratio of one solver's runtime to
the best runtime appears in \cite{SCB95a}, with solvers rated by
the percentage of problems for which a solver's time is termed \textit{very
competitive} or \textit{competitive}.
The ratio approach avoids most of the difficulties that we
have discussed, providing information on the percent improvement and 
eliminating the negative effects of allowing a small
portion of the problems to dominate the conclusions.
The main disadvantage of this approach lies in the author's
arbitrary choice of limits to define the borders of
\textit{very competitive} and \textit{competitive}.

In Section \ref{perf}, we 
introduce \textit{performance profiles} as a tool 
for evaluating and comparing the performance of 
optimization software. The
performance profile for a solver is the (cumulative) distribution
function for a performance metric. In this paper we
use the ratio of the computing time of the solver 
versus the best time of all of the solvers as the performance metric. 
Section \ref{data}
provides an analysis of the test set and solvers used in the benchmark
results of  Sections \ref{case1} and \ref{case2}.
This analysis is necessary to understand the limitations of
the benchmarking process.

Sections \ref{case1} and \ref{case2} demonstrate the use
of performance profiles with results \cite {EDD00}
obtained with version 2.0 of
the \cops\ \cite{cops-home} test set.
We show that performance profiles
eliminate the influence of a small number of problems
on the benchmarking process and
the sensitivity of results associated
with the ranking of solvers. 
Performance profiles provide a means of visualizing the
expected performance difference among many solvers, while avoiding 
arbitrary parameter choices and the need to discard solver
failures from the performance data.

We conclude in Section \ref{lp}
by showing how performance profiles apply to
the data \cite {Benchmarks} of Mittelmann for 
linear programming solvers.
This section provides another case study of the use of performance
profiles and also shows that performance profiles can be
applied to a wide range of performance data.
\section{Performance Evaluation}
\label{perf}
Benchmark results are 
generated by running a solver on a set $ \cP $ of problems and
recording information of interest such as the number of
function evaluations and the computing time.
In this section we introduce the notion of a performance profile
as a means to evaluate and compare the performance of the set of 
solvers $ \cS $ on a test set $ \cP $.

We assume that we have $ n_s $ solvers and $ n_p $ problems.
We are interested in using computing time as a performance measure;
although, the ideas below can be used with other measures.
For each problem $p$ and solver $s$, we define
\[
t_{p,s} = \mbox{computing time required to solve problem $p$ by solver $s$.}
\]
If, for example, the number of function evaluations is the performance
measure of interest, set $ t_{p,s} $ accordingly.  

We require a baseline for comparisons.
We compare the performance on problem $p$ by solver $s$
with the best performance by any solver on this problem; that is,
we use the \textit{performance ratio}
\[
r_{p,s} = \frac{t_{p,s}}{\min \{ t_{p,s} : s \in \cS \}} .
\]
We assume that a parameter $ r_M \ge r_{p,s}$ for all $p,s$ 
is chosen, and $ r_{p,s} = r_M $ if and only if solver $s$ does not solve
problem $p$.  We will show that the choice of $ r_M $ does not affect 
the performance evaluation.

The performance of solver $s$ on any given problem may be of interest,
but we would like to obtain an overall assessment of the performance
of the solver. If we define
\[
\rho_s ( \tau ) = \frac{1}{n_p} 
               \mbox{size} \Bigl \{ p \in \cP : r_{p,s} \le \tau  \Bigr \},
\]
then $ \rho_s ( \tau ) $ is the probability for solver $ s \in \cS $
that a performance ratio $ r_{p,s} $ 
is within a factor $ \tau \in \R $ of the best possible ratio.  The
function $ \rho_s $ is the (cumulative) distribution function for the
performance ratio.

We use the term \textit{performance profile} for the 
distribution function of a performance metric.
Our claim is that a plot of the performance profile
reveals all of the major performance characteristics.
In particular, if the set of problems $ \cP $ is suitably large and 
representative of problems that are likely to occur in applications, then
solvers with large probability $ \rho_s ( \tau ) $ are
to be preferred.

The term performance profile has also been used for a plot of some
performance metric versus a problem parameter. For example,
Higham \cite [pages 296--297] {NJH96} plots the ratio
$ \gamma/\| A \|_1 $, where $ \gamma $ is the estimate for the
$ l_1 $ norm of a matrix $A$ produced by the LAPACK condition number estimator.
Note that in Higham's use of the term performance profile
there is no attempt at determining a distribution function.

The performance profile $ \rho_s : \R \mapsto [0,1 ] $
for a solver is a nondecreasing, piecewise constant
function, continuous from the right at each breakpoint.
The value of $ \rho_s (1) $ is the probability that
the solver will win over the rest of the solvers.
Thus, if we are interested only in the number of wins, we need only
to compare the values of $ \rho_s (1) $ for all of the solvers.  

The definition of the performance profile for large values requires
some care. We assume that $ r_{p,s} \in [ 1, r_M ] $ and that
$ r_{p,s} = r_M $ only when problem $p$ is not
solved by solver $s$. 
As a result of this convention, $ \rho_s (r_M) = 1 $, and 
\[
\rho_s^* \equiv \lim_{\tau \to r_M^- } \rho_s ( \tau )
\]
is the probability that the solver solves a problem. 
Thus, if we are interested only in solvers with
a high probability of success, then we
need to compare the values of $ \rho_s^* $ for all solvers
and choose the solvers with the largest value.
The value of $ \rho_s^* $ can be readily seen in a performance 
profile because $ \rho_s $ flatlines for large values of $ \tau $; that is, 
$ \rho_s ( \tau ) = \rho_s^* $ for 
$ \tau \in  [ r_S, r_M ) $ for some
$ r_S \lt r_M $.

%

An important property of performance
profiles is that they are insensitive to the results on a small
number of problems. This claim is based on the observation
that if $ \rho_s $ and  $ \hat \rho_s $ are
defined, respectively, by the observed 
time sets $ t_{p,s} $ and $ \hat t_{p,s} $, where
\[
\hat t_{p,s} = t_{p,s}, \qquad p \in \cP\setminus\{q\},
\]
for some problem $q \in \cP$, then
$ \hat r_{p,s} = r_{p,s} $ for $ p \in \cP\setminus\{q\} $.
Since only the ratio $ \hat r_{q,s} $ changes for any $ s \in \cS $,
\[
| \rho_s(\tau) - \hat \rho_s(\tau)| \leq \frac {1} {n_p}, \qquad \tau \in \R,
\]
for $ s\in\cS $.
Moreover, $ \hat \rho_s ( \tau ) = \rho_s ( \tau ) $ for
$ \tau < \min\{r_{q,s}, \hat r_{q,s} \}$ or 
$ \tau \geq \max\{r_{q,s}, \hat r_{q,s} \}$.
Thus, if $ n_p $ is reasonably large, then the result on a particular
problem $q$ does not greatly affect the performance profiles $ \rho_s $.

Not only are performance profiles relatively insensitive to changes in 
results on a small number of problems, they are also largely unaffected by 
small changes in results over many problems.  We demonstrate this property
by showing that small changes from $r_{p,s}$ to $\hat r_{p,s}$ 
result in a correspondingly small
$L_1$ error between $\rho_s$ and $\hat \rho_s$.

\begin {theorem}
Let $ r_i $ and $ \hat r_i $ for $ 1 \le i \le n_p $ be performance ratios
for some solver, and let
$ \rho $ and $ \hat \rho $ be, respectively, the performance profiles
defined by these ratios.
If 
\begin{equation}
\label{eq:bound1}
| r_i - \hat r_i | \le \epsilon , \qquad 1 \le i \le n_p 
\end{equation}
for some $ \epsilon > 0 $, then
\[
\int _{1}^{\infty} 
\left | \rho ( t ) - \hat \rho (t) \right | \, dt \le \epsilon
\]
\end {theorem}

\proof
Since performance profiles do not depend on the ordering
of the data, we can assume that $ \{ r_i \} $ is
monotonically increasing. We can reorder 
the sequence  $ \{ \hat r_i \} $ so that
it is also monotonically increasing, and \Ref{eq:bound1} still holds.
These reorderings guarantee
that $ \rho ( t ) = i/n_p $ for
$ t \in [ r_i , r_{i+1} ) $, with a similar result for
$ \hat \rho $.
We now show that for any integer $k$ with $ 1 \le k \le n_p $,
\begin{equation}
\label{eq:est1}
\int _{1}^{s_{k}}
\left | \rho ( t ) - \hat \rho (t) \right | \, dt \le 
 k \left ( \frac {\epsilon}{n_p} \right ) ,
\end{equation}
where  $ s_k = \max ( r_k , \hat r_k ) $, and
\begin{equation}
\label{eq:bound2}
| r_i - \hat r_i | \le \epsilon , \quad 1 \le i \le k, 
\quad \mbox{and} \quad
\hat r_i = r_i , \quad k < i \le n_p . 
\end{equation}
The proof is completed when $k=n_p$.

The case $ k = 1 $ follows directly from
the definition of a performance profile, so assume that
\Ref{eq:est1} holds for any performance data such that
\Ref{eq:bound2} holds.
We now prove that \Ref{eq:est1} holds for $k+1$ by proving that 
\begin{equation}
\label{eq:bound3}
\int _{s_k}^{s_{k+1}}
\left | \rho ( t ) - \hat \rho (t) \right | \, dt \le 
\left ( \frac {\epsilon}{n_p} \right ), \qquad k < n_p.
\end{equation}
Together, \Ref{eq:est1} and \Ref{eq:bound3} show that \Ref{eq:est1}
holds for $k+1$.

We present the proof for the case when  $ \hat r_k \le r_k $.
A similar argument can be made for $r_k \leq \hat r_k$.
If $ \hat r_k \le r_k $ then $ s_k = r_k $ and $ \hat r_k \le r_k \le r_{k+1} $.
The argument depends on the position of $ \hat r _{k+1} $
and makes repeated use of the fact that $ \rho ( t ) = k/n_p $ for
$ t \in [ r_k , r_{k+1} ) $, with a similar result for
$ \hat \rho $.

If $ r_{k+1} \le \hat r_{k+1} $ then 
$ \rho(t) = \hat \rho (t) $ in $ [ r_k , r_{k+1} ) $.
Also note that
$ \left | \rho ( t ) - \hat \rho (t) \right | = 1/n_p $ 
in $ [ r_{k+1} , \hat r_{k+1} ) $.
Hence, \Ref{eq:bound3} holds with $ s_{k+1} = \hat r_{k+1} $.

The case where $ \hat r_{k+1} \le  r_{k+1} $ makes use
of the observation that
$ \hat r_i = r_i \ge r_{k+1} $ for $ i > k + 1 $.
If $ r_k \le \hat r_{k+1} \le r_{k+1} $, then
$ \rho(t) = \hat \rho (t) $ in $ [ r_k , \hat r_{k+1} ) $,
and
$ \left | \rho ( t ) - \hat \rho (t) \right | = 1/n_p $ in
$ [ \hat r_{k+1} , r_{k+1} ) $.
Hence, \Ref{eq:bound3} holds.
On the other hand, if
$ \hat r_k \le \hat r_{k+1} \le r_{k} $, then we only need to note
that
$ \left | \rho ( t ) - \hat \rho (t) \right | = 1/n_p $ in
$ [  r_{k} , r_{k+1} ) $ in order to conclude that
\Ref{eq:bound3} holds. 

We have shown that \Ref{eq:est1} holds for all integers $k$ with
$ 1 \le k \le n_p $. In particular, the case $ k = n_p $ yields
our result since  $ \rho(t) = \hat \rho (t) $ for 
$ t \in [ s_{n_p} , \infty) $.
\qed

\section{Benchmarking Data}

\label{data}

The timing data used to compute the
performance profiles in Sections \ref{case1} and \ref{case2}
are generated with the \cops\ test set,
which currently consists of seventeen different applications, 
all models in the AMPL \cite{AMPL} modeling
language.  The choice of the test problem set $ \cP $ is always a source
of disagreement because there is no consensus on how
to choose problems.
The \cops\ problems are selected to be interesting and difficult,
but these criteria are subjective.
For each of the applications in the \cops\ set we use
four instances of the application obtained by varying
a parameter in the application, for example, the number
of grid points in a discretization.  Application descriptions
and complete absolute timing results for the full test set
are given in \cite{EDD00}.

Section \ref{case1} focuses on only the subset of the eleven optimal control and
parameter estimation applications
in the \cops\ set, while the discussion in Section \ref{case2} covers the 
complete performance results.  Accordingly, we provide here information
specific to this subset of the \cops\ problems as
well as an analysis of the test set as a whole.
Table~\ref{table:cops-stats} gives the quartiles
for three problem parameters: the number of
variables $n$, the number of constraints, and the 
ratio $ (n -n_e)/n $, where $ n_e $ is the number
of equality constraints. In the optimization literature,
$ n - n_e $ is often called the degrees of freedom of
the problem, since it is an upper bound on the number
of variables that are free at the solution.

The data in Table 
\ref{table:cops-stats} 
is fairly representative of the distribution of these 
parameters throughout the test set and shows
that at least three-fourths of the problems have
the number of variables $n$ in the interval $ [400, 5000 ] $.
Our aim was to avoid problems where $n$ was 
in the range $ [ 1 , 50 ] $ because other benchmarking
problem sets tend to 
have a preponderance of problems with $ n $ in this range.
The main difference between the full \cops\ set and the
\cops\ subset is that the \cops\ subset is more
constrained with $ n_e \ge  n/2 $ for all the problems.
Another feature of the \cops\ subset is that the equality
constraints are the result of either difference or collocation
approximations to differential equations.

We ran our final complete runs with the
same options for all models.  The options involve setting the
output level for each solver so that we can gather the data 
we need, increasing the iteration limits as much as allowed, and 
increasing the super-basics limits for MINOS and SNOPT to 5000.  
None of the failures we record in the final trials include any 
solver error messages about having violated these limits.
While we relieved restrictions unnecessary for our testing, all
other parameters were left at the solvers' default settings.

\begin{table}
\small
\caption{Problem data for \cops\ test set}
\begin{tabular}{@{}l c c c c c|c c c c c@{}}
\vspace{-7pt}
& \multicolumn{5}{c|}{Full \cops} & \multicolumn{5}{c@{}}{\cops\ subset} \\ 
\vspace{-3pt}
& \multicolumn{5}{r@{}|}{\hrulefill} &  \multicolumn{5}{@{}l@{}}{\hrulefill} \\
&  $\min$  &  $q_1$  &  $q_2$  &  $q_3$  &  $\max$ 
&  $\min$  &  $q_1$  &  $q_2$  &  $q_3$  &  $\max$ \\ \hline 
\vspace{1pt}
Num. variables & 48 & 400 & 1000 & 2402 & 5000 & 100 & 449 & 899 & 2000 & 4815 \\
Num. constraints & 0 & 150 & 498 & 1598 & 5048 & 51 & 400 & 800 & 1601 & 4797 \\
Degrees freedom & 0 & 23 & 148 & 401 & 5000 & 0 & 5 & 99 & 201 & 1198 \\
Deg. freedom (\%) & 0.0 & 1.0 & 33.2 & 100.0 & 100.0 &
0.0 & 0.4 & 19.8 & 33.1 & 49.9\\
\hline
\end{tabular}  
\label{table:cops-stats}
\end{table}

The script for generating the timing data
sends a problem to each solver successively, 
so as to minimize the effect of
fluctuation in the machine load.  The script tracks the wall-clock
time from the start of the solve, killing any process that runs 3,600
seconds, which we declare unsuccessful, and beginning the next solve.
We cycle through all the problems, recording the
wall-clock time as well as the combination of AMPL system time (to
interpret the model and compute varying amounts of derivative
information required by each solver) and AMPL solver time for each 
model variation.  
We repeat the cycle for any model for
which one of the solvers' AMPL time and the wall-clock time differ 
by more than 10 percent.  To further ensure consistency, we have 
verified that the 
AMPL time results we present could be reproduced to within 10 percent 
accuracy.  All computations were done on a SparcULTRA2
running Solaris~7.

We have ignored the effects of the stopping criteria and the memory
requirements of the solvers.
Ideally we would have used the same stopping criteria, but this
is not possible in the AMPL environment. In any case, differences in
computing time due to the stopping criteria are not likely to change
computing times by more than a factor of two.
Memory requirements can also play an important role.
In particular, solvers that use direct linear equation solvers are often more
efficient in terms of computing time provided there is enough memory.

The solvers that we benchmark have
different requirements.  MINOS and SNOPT use only first-order information, 
while LANCELOT and LOQO need second-order information.
The use of second-order information can reduce the number of
iterations, but the cost per iteration usually increases.  In
addition, obtaining second-order information is more costly and
may not even be possible.  
MINOS and SNOPT are specifically designed for problems with a modest
number of degrees of freedom, while this is not the case for LANCELOT
and LOQO.
As an example of comparing solvers with similar requirements,
Section \ref{lp} shows the performance of linear programming solvers.

\section{Case Study: Optimal Control and Parameter Estimation Problems}

\label{case1}
\begin{figure*}[bh]
\centerline {\includegraphics[height=9.3cm]{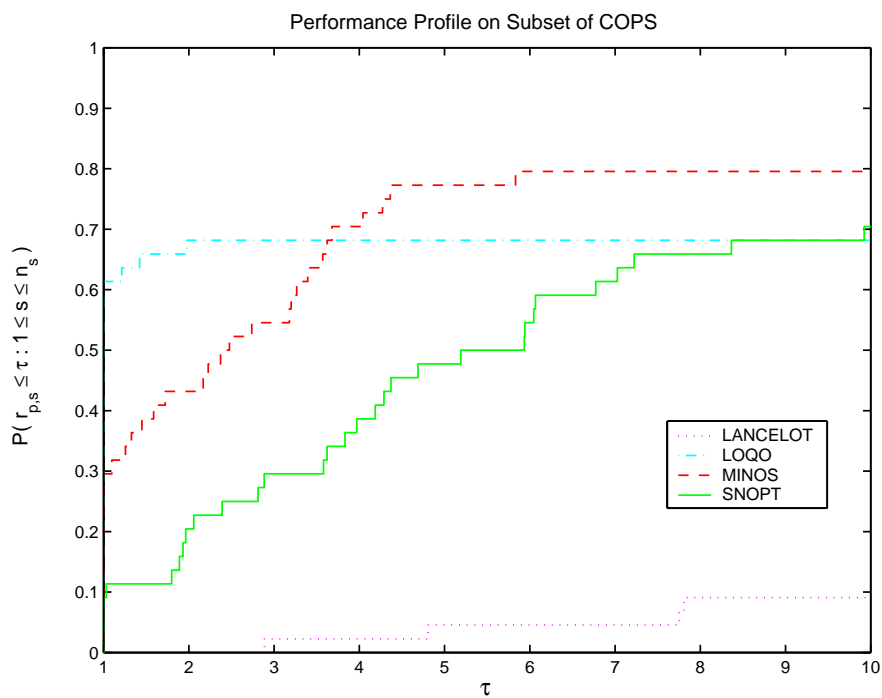}}
\caption {Performance profile on $ [0, 10 ] $}
\label {figure:perf10}
\end{figure*}

\begin{figure*}[tb]
\centerline {\includegraphics[height=9.3cm]{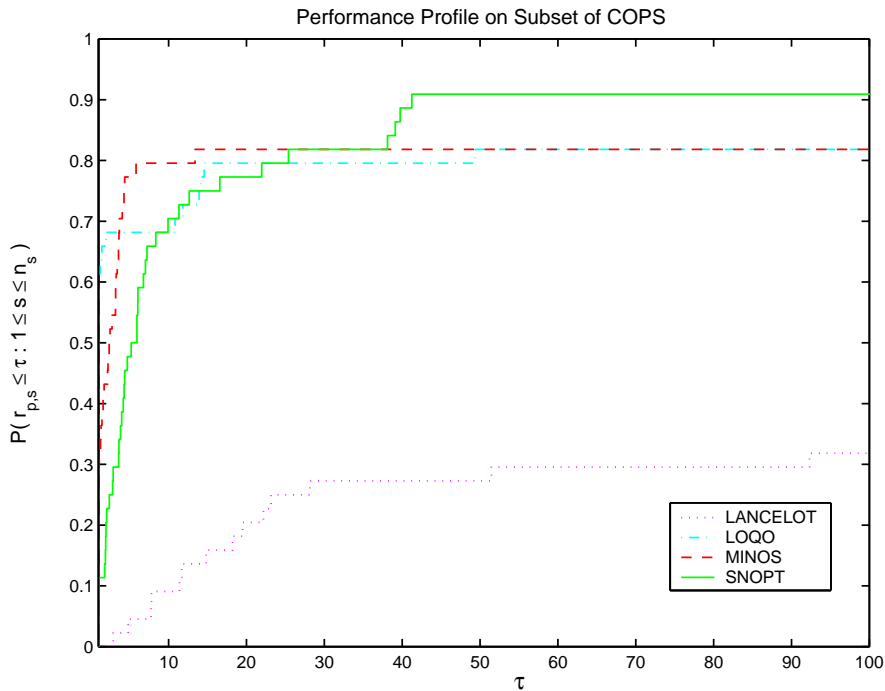}}
\caption {Performance profile on $ [0, 100 ] $}
\label {figure:perf100}
\end{figure*}

We now examine the performance of 
LANCELOT~\cite{ARC92}, MINOS~\cite{BAM95}, SNOPT~\cite{PEG97}, and 
LOQO~\cite{RJV97b} on the subset of the
optimal control and parameter estimation
problems in the \cops\ \cite{cops-home} test set.
Figures \ref{figure:perf10} and \ref{figure:perf100}
show the performance profiles in different ranges to
reflect various areas of interest.  
Our purpose is to show how performance profiles provide 
objective information for analysis of a large test set.
Figure \ref{figure:perf10} shows the performance profiles of
the four solvers for small values of $ \tau $.
By showing the ratios of solver times, we eliminate any weight of 
importance that taking straight time differences might give 
to the problems that require a long run time of every solver.
We find no need to eliminate any
test problems from discussion.  For this reason, solvers 
receive their due credit for completing
problems for which one or more of the other solvers fails.
In particular,
$1 - \rho_s(\tau)$ is the fraction
of problems that the solver cannot solve within a factor $\tau$ of
the best solver, including problems for which the solver
in question fails.

From this figure it is clear that LOQO has the
most wins (has the highest probability of being the
optimal solver) and that the probability that LOQO is the winner on 
a given problem is about $ .61 $.
If we choose being within a factor of $4$ of the best solver as the scope
of our interest, then either LOQO or MINOS would suffice; but 
the performance profile shows that the probability that these two 
solvers can solve a job within a factor $4$ of the best solver is only 
about $ 70 \% $.
SNOPT has a lower number of wins than either LOQO or MINOS, but its 
performance becomes much more competitive if we extend our $\tau$ of 
interest to $7$.

Figure \ref{figure:perf100} shows the performance profiles for all the
solvers in the interval $ [ 1, 100 ] $.  If we are interested in the solver
that can solve $75\%$ of the problems with the greatest efficiency,
then MINOS stands out.  If we hold to more stringent probabilities of 
completing a solve successfully, then SNOPT captures our attention with its
ability to solve over $90\%$ of this \cops\ subset, as
displayed by the height of its performance profile for $\tau > 40$.
This graph displays the potential for large discrepancies in the 
performance ratios on a substantial percentage of the problems. 
Another point of interest is that LOQO,
MINOS, and SNOPT each have the best probability $\rho_s(\tau)$
for $\tau$ in some interval, with similar performance
in the interval $ [15,40] $.

An observation that emerges from these
figures is the lack of consistency in quartile values of time ratios.
The top three solvers share a minimum ratio of $1$, and 
LOQO and MINOS also share first quartile values of $1$.  
In other words, these two solvers are the best solvers on at
least 25\% of the problems. 
LOQO bests MINOS's median value with 
$1$ compared with $2.4$, but MINOS comes back with a third quartile 
ratio of $4.3$ versus $13.9$ for LOQO, with SNOPT mixing results
further by also beating LOQO with $12.6$.
By looking at Figures \ref{figure:perf10} and \ref{figure:perf100}, we
see that progress between quartiles does not necessarily 
proceed linearly; hence, we really lose information if we do not
provide the full data.
Also, the maximum ratio would be $r_M$ for our testing, and no
obvious alternative value exists.  As an alternative to providing
only quartile values, however, the performance profile yields 
much more information about a solver's strengths and weaknesses.

\begin{figure*}
\centerline {\includegraphics[height=9.3cm]{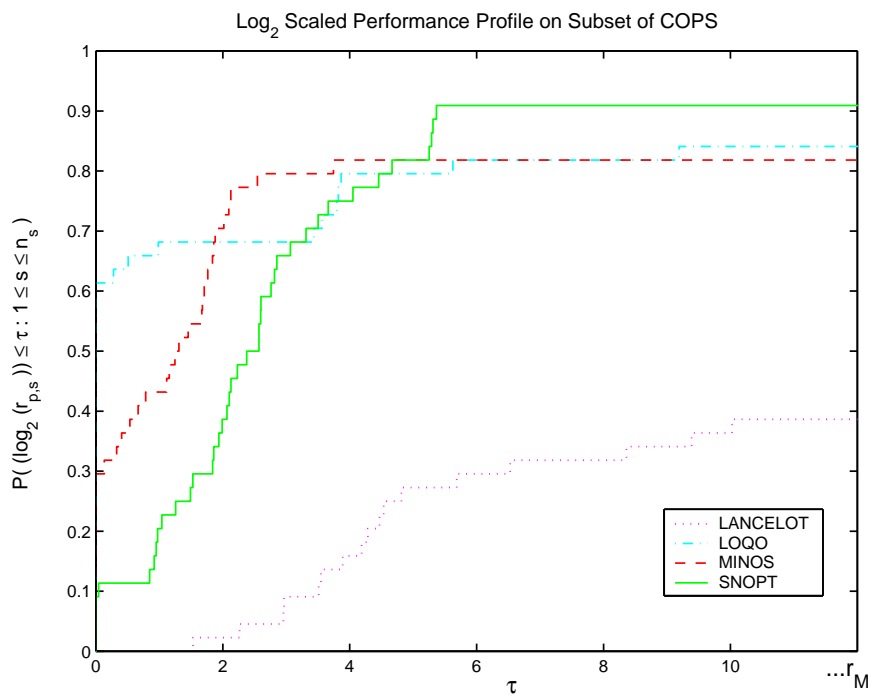}}
\caption {Performance profile in a $ \log_{2} $ scale}
\label {figure:log2}
\end{figure*}

We have seen that at least two graphs may be needed to examine the performance
of the solvers. Even extending $\tau$ to $100$, we fail to capture the
complete performance data for LANCELOT and LOQO.
As a final option, we display a log scale of
the performance profiles.  In this way, we can show all activity that
takes place with $\tau < r_M$ and grasp the full implications of 
our test data regarding the solvers' probability of successfully 
handling a problem.
Since we are also interested in the
behavior for $ \tau $ close to unity, we use a base of 2 for the
scale. In other words, we plot
\[
\tau \mapsto \frac{1}{n_p} \mbox{size} \left \{ p \in \cP : 
\log_2 \left ( r_{p,s} \right ) \le \tau  \right \}
\]
in Figure \ref{figure:log2}.  This graph reveals all the features of the
previous two graphs and thus captures the performance of all the solvers.
The disadvantage is that the interpretation of the
graph is not as intuitive, since we are using a log scale.

Figures \ref{figure:perf10} and 
\ref{figure:perf100} are mapped into a new scale to reflect all data, 
requiring 
at least the interval $ [0 ,  \log_2 (1043) ] $
in Figure \ref{figure:log2} to include the largest $r_{p,s} < r_M$.  
We extend the range slightly to show the
flatlining of all solvers.
The new figure contains
all the information of the other two figures and, in addition,
shows that each of the solvers fails on at least $ 8\% $
of the problems. This is not an unreasonable performance for the
\cops\ test set because these problems were generally chosen
to be difficult.

\section{Case Study: The Full COPS}

\label{case2}
\begin{figure*}
\centerline {\includegraphics[height=9.3cm]{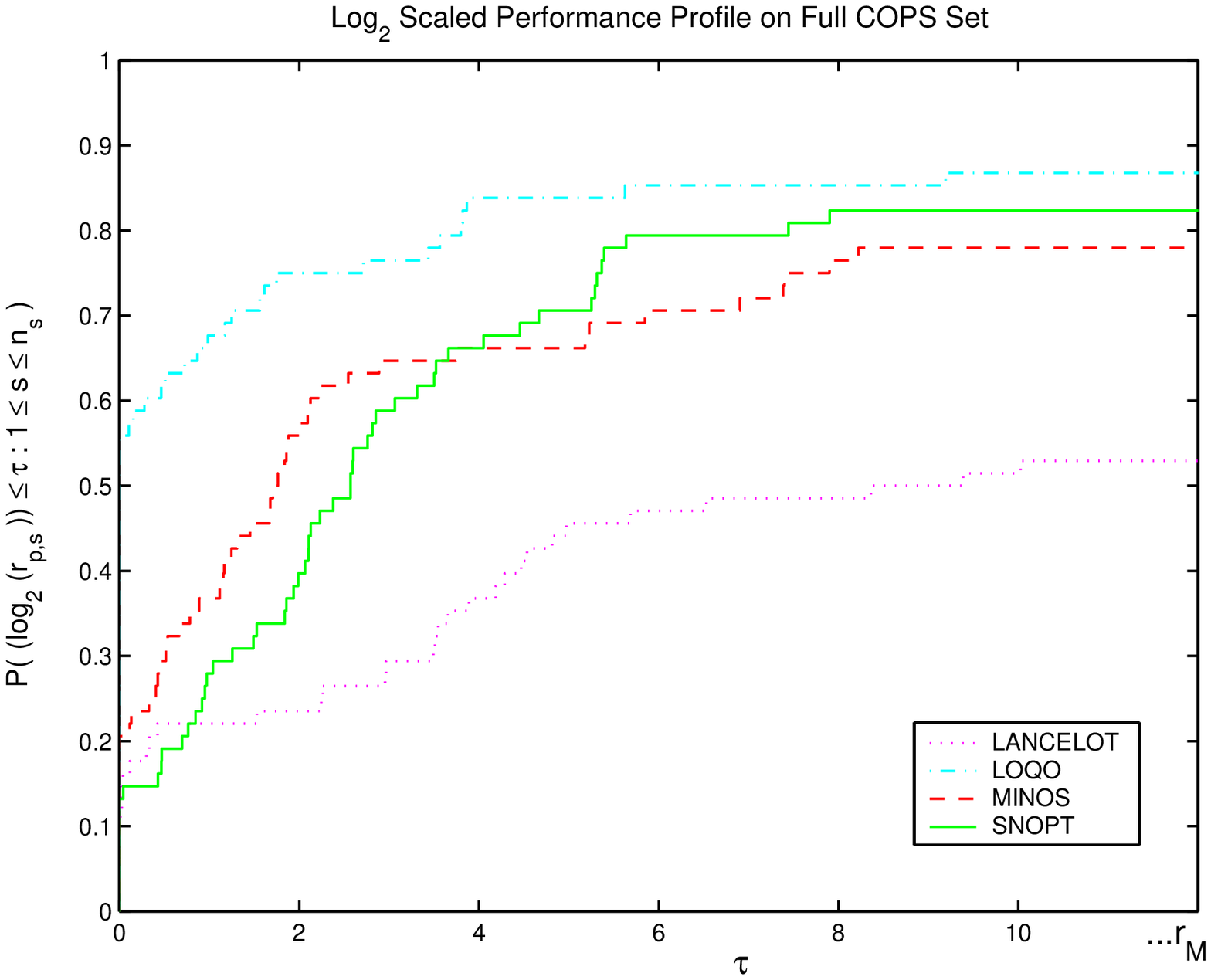}}
\caption {Performance profile for full COPS set}
\label {figure:copsprof}
\end{figure*}
We now expand our analysis of the data to include all
the problems in version 2.0 of the \cops\ \cite{cops-home}
test set. We present in
Figure \ref{figure:copsprof} 
a log$_2$ scaled view of the performance profiles for the 
solvers on that test set.

Figure \ref{figure:copsprof} gives a clear indication of the
relative performance of each solver. 
As in the performance profiles
in Section \ref{case1}, this figure shows that performance profiles
eliminate the undue influence of a small number of problems
on the benchmarking process and
the sensitivity of the results associated
with the ranking of solvers. In addition,
performance profiles provide an estimate of the
expected performance difference between solvers.

The most significant aspect of Figure  \ref{figure:copsprof},
as compared with Figure \ref{figure:log2},
is that on this test set LOQO dominates all other solvers:
the performance profile for LOQO lies above
all others for all performance ratios.
The interpretation of the results  in Figure~\ref{figure:copsprof}
is important. In particular, these results
do not imply that LOQO is faster on every problem.
They indicate only that, for any $ \tau \ge 1 $,
LOQO solves more problems within a factor of $ \tau $ of any other 
solver time.  Moreover, by examining $\rho_s (1)$ and $\rho_s(r_M)$,
we can also say that LOQO is the fastest solver on approximately 58\% of
the problems, and that LOQO solves the most problems (about 87\%)
to optimality.

The difference between the results in Section \ref{case1}
and these results is due to a number of factors.
First of all, as can be seen in Table \ref{table:cops-stats},
the degrees of freedom for
the full \cops\ test set is much larger than for the
restricted subset of optimal control and parameter estimation problems.
Since, as noted in Section \ref{data},
MINOS and SNOPT are designed for problems with a modest
number of degrees of freedom, we should expect the
performance of MINOS and SNOPT to deteriorate on the
full \cops\ set.
This deterioration can be seen by comparing
Figure  \ref{figure:copsprof} with Figure \ref{figure:log2}
and noting that the performance profiles of MINOS and SNOPT
are similar but generally lower in Figure~\ref{figure:copsprof}.

Another reason for the 
difference between the results in Section \ref{case1}
and these results is that
MINOS and SNOPT use only first-order information,
while LOQO uses second-order information.
The benefit of using second-order information usually 
increases as the number of variables increases,
so this is another factor that benefits LOQO.

The results in this section underscore our observation
that performance profiles provide a convenient tool
for comparing and evaluating the performance of optimization solvers,
but, like all tools, performance profiles must be used with care.
A performance profile
reflects the performance only on the data being used, and thus it
is important to understand the test set and the solvers
used in the benchmark.

\section{Case Study: Linear Programming}

\label{lp}

\begin{figure*}
\centerline {\includegraphics[height=9.5cm]{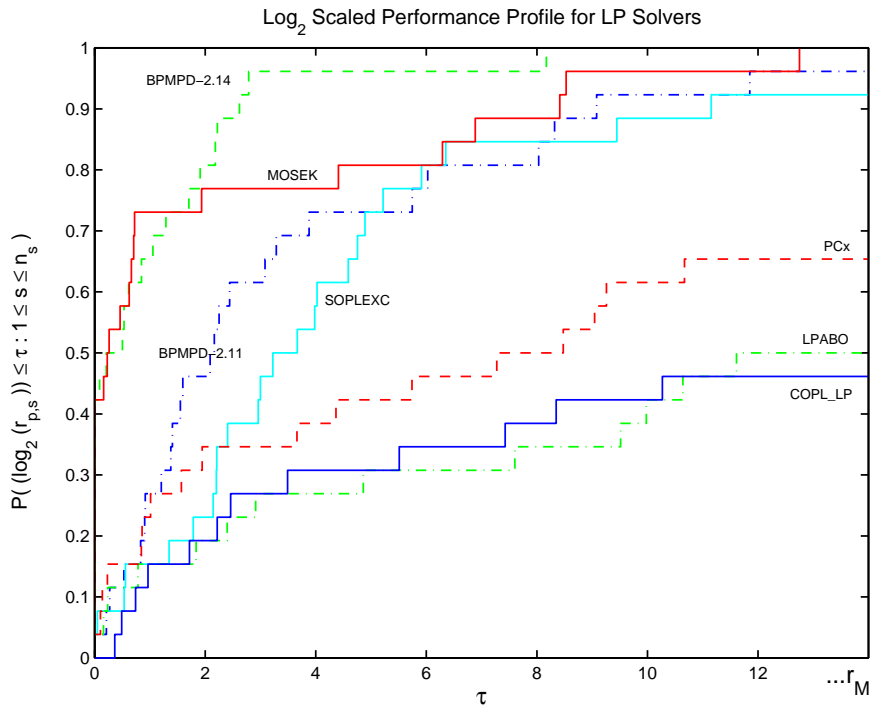}}
\caption {Performance profile for linear programming solvers}
\label {figure:HANSlog}
\end{figure*}

Performance profiles can be used
to compare the performance of two solvers, but performance
profiles are most useful in comparing several solvers. 
Because large amounts of data are generated in these situations,
trends in performance are often difficult to see.
As a case study, we use data obtained
by Mittelmann \cite{Benchmarks}.
Figure \ref{figure:HANSlog} shows a plot of the performance
profile for the time ratios in the data
\textit{Benchmark of LP solvers on a Linux-PC} (5-25-2000),
which includes results for
COPL\_LP (1.0), PCx (1.1),
SOPLEX (1.1), LPABO (5.6), MOSEK (1.0b), BPMPD (2.11), 
and BPMPD (2.14).  

In keeping with our graphing practices
with the \cops\ set, we designate as failures those 
solves that are marked in the original table as stopping
close to the final solution without convergence under the solver's
stopping criteria.   One feature we see in the graph of Mittelmann's
results that does not appear in the \cops\ graphs is the visual
display of solvers that never flatline.  In other words, the solvers
that climb off the graph are those that solve all of the test
problems successfully.  As with Figure \ref{figure:log2}, all of the
events in the data fit into this log-scaled representation.  While
this data set cannot be universally representative of benchmarking
results by any means, it does show that our reporting technique is
applicable beyond our own results.

As in the case studies in Sections \ref{case1} and
\ref{case2}, the results in Figure \ref{figure:HANSlog} give an
indication of the performance of LP solvers only on the data set used
to generate these results. In particular, the test set
used to generate Figure \ref{figure:HANSlog} includes only
problems selected by Mittelmann for his benchmark. The advantage of these
results is that, unlike the solvers in Sections \ref{case1} and
\ref{case2}, all solvers in Figure \ref{figure:HANSlog}
have the same requirements.

\section{Conclusions}
We have shown that performance profiles combine the best features
of other tools for benchmarking and comparing optimization solvers.
Clearly, the use of performance profiles
is not restricted to optimization solvers
and can be used to compare solvers in other areas.

We have not addressed the issue of how to select a collection
of test problems to justify performance claims.
Instead, we have provided a tool -- performance profiles --
for evaluating the performance of two or more solvers on
a given set of test problems.
If the data is obtained  by following careful guidelines
\cite{Crowder:1979:RCE,JacBNP91}, then performance profiles can be used to
justify performance claims.
We emphasize that claims on the relative performance of the
solvers on problems not in the test set should be made with care.

The Perl script \texttt{perf.pl} on the \cops\ site \cite{cops-home}
generates performance profiles formatted as Matlab commands to produce 
a composite graph as in Figures \ref{figure:perf10} and 
\ref{figure:perf100}.
The script accepts data for several solvers and plots 
the performance profile on
an interval calculated to show the full area of activity.
The area displayed and scale of the graph can then be adjusted within Matlab
to reflect particular benchmarking interests.

\begin{acknowledgement}

\label{acks}

This work was supported by the Mathematical, Information, and
Computational Sciences Division subprogram of the Office of Advanced
Scientific Computing, U.S. Department of Energy, under Contract
W-31-109-Eng-38, and
by the National Science Foundation 
(Challenges in Computational Science) grant CDA-9726385
and (Information Technology Research) grant CCR-0082807. 

Version 1.0 of the \cops\ problems was developed by
Alexander Bondarenko, David Bortz, Liz Dolan, and Michael Merritt.
Their contributions were essential because,
in many cases, version 2.0 of the problems
are closely related to the original version.

Alex Bondarenko, Nick Gould, Sven Leyffer and Bob Vanderbei contributed
interesting and spirited
discussions on problem formulation, while
Bob Fourer and David Gay generously shared their
AMPL expertise with us.
Hans Mittelmann deserves special note
for paving the way for \cops\ with
his benchmarking work.

We also thank the referees and the associate editor, Jorge Nocedal, for
their comments. One of the referees provided a thoughtful report
that led to additional details on the sensitivity of performance profiles.
\end{acknowledgement}
\addcontentsline{toc}{section}{Acknowledgments}
\bibliographystyle{}
\bibliography{}
\addcontentsline{toc}{section}{References}


\end{document}